\documentclass[a4paper,11pt]{article}
\usepackage{pos}
\usepackage{graphicx} % Required for inserting images

\title{Hadronic vacuum polarization contribution to the muon g-2 on Euclidean windows from tau data}

\author[a,b]{Pere Masjuan}
\author*[b]{Alejandro Miranda}
\author[c]{Pablo Roig}

\affiliation[a]{Grup de Física Teòrica, Departament de Física, Universitat Autònoma de Barcelona, 08193
Bellaterra (Barcelona), Spain.}
\affiliation[b]{Institut de Física d’Altes Energies (IFAE) and The Barcelona Institute of Science and
Technology (BIST), Campus UAB, 08193 Bellaterra (Barcelona), Spain.}

\affiliation[c]{Departamento de Física, Centro de Investigación y de Estudios Avanzados del Instituto
Politécnico Nacional, Apdo. Postal 14-740, 07000 Ciudad de México, México.}

\emailAdd{masjuan@ifae.es}
\emailAdd{jmiranda@ifae.es}
\emailAdd{pablo.roig@cinvestav.mx}

\abstract{We computed for the first time the $\tau$ data-driven Euclidean windows for the hadronic vacuum polarization contribution to the muon g-2. We showed that $\tau$-based results agree with the available lattice window evaluations and with the full result. On the intermediate window, where all lattice evaluations are rather precise and agree, $\tau$-based results are compatible with them. This is particularly interesting, given that the disagreement of the $e^+e^-$ data-driven result with the lattice values in this window is the main cause for their discrepancy, affecting the interpretation of the $a_\mu$ measurement in terms of possible new physics.}

\FullConference{10th International Conference on Quarks and Nuclear Physics (QNP2024)\\
8-12 July, 2024\\
Barcelona, Spain\\}

\date{}
\begin{document}

\maketitle

\section{Introduction: \texorpdfstring{$\boldsymbol{a_\mu}$}{Lg} and HVP}
The uncertainty of the SM prediction for the anomalous magnetic moment of the muon, $a_\mu$, is dominated by the hadronic contributions which, according to the White Paper \cite{Aoyama:2020ynm} (to be updated by the end of this year), are at the level of 0.37(0.15) ppm in the Hadronic Vacuum Polarization, HVP (Hadronic Light-by-light) pieces, respectively. This uncertainty of $44\times10^{-11}$ is slightly more than twice that of the experimental average, coming from the BNL \cite{Muong-2:2006rrc} and FNAL \cite{Muong-2:2021ojo,Muong-2:2023cdq} measurements, that is 0.19 ppm, close to the FNAL goal of 0.14 ppm. The E34 experiment at J-PARC aims to perform ultra-precise measurements of $a_\mu$ and the muon electric dipole moment using a different method, which will have a completely independent systematic uncertainty \cite{Iinuma:2011zz}. Taken at face value, the difference between the White Paper results and the $a_\mu$ average measurement would be slightly larger than $5\sigma$, the usual reference to claim (indirect) new physics discovery.

The SM uncertainty is thus dominated by the HVP contribution, which has been traditionally computed via a dispersion relation in terms of experimental data \cite{Gourdin:1969dm}
\begin{equation}\begin{split}
a_\mu^\mathrm{HVP,LO}&=\frac{\alpha^2}{3\pi^3}\int_{m_\pi^2}^{\infty}ds \frac{K(s)}{s}R(s)\,,\\
R(s)&=\frac{\sigma^0(e^+e^- \to\mathrm{hadrons}(\gamma))}{\sigma_{pt}}\,,\qquad \sigma_{pt}=\frac{4\pi\alpha^2}{3s}\,,
\end{split}\end{equation}
where $K(s)$ is a QED kernel function concentrated at low energies~\cite{Brodsky:1967sr}. %with the kernel $K(s)\sim1/s$.
Alternatively, it can also be obtained using the spectral function of $\tau^-\to\nu_\tau\,\mathrm{hadron}$ decays, which can be related to the isovector component of the required $e^+e^-$ cross-section through an isospin rotation. Given the current puzzle with $e^+e^-$ data (that we will summarize shortly), we insist it is a good strategy to keep using both. Particularly, tau data can be very useful for the $\pi\pi$ contribution, which amounts to $\sim73\%$ of the whole $a_\mu^\mathrm{HVP,LO}$ and to $\sim58\%$ of its uncertainty at low energies \cite{Jegerlehner:2017gek}. For this final state ($\sqrt{s}$ is the $\pi\pi$ invariant mass),
\begin{equation}\begin{split}
\sigma_{\pi^+\pi^-(\gamma)}^0(s)&=\frac{\pi\alpha^2\beta^3_{\pi^+\pi^-}(s)}{3s}|F_V(s)|^2\,\\&=\,\frac{K_\sigma(s)}{K_\Gamma(s)}\frac{d\Gamma_{\pi\pi[\gamma]}}{ds}\frac{R_\mathrm{IB}(s)}{S_\mathrm{EW}}\,,
\end{split}\end{equation}
where the first factor of the last member includes kinematical functions, the second one is the tau spectra, and short-distance electroweak radiative corrections are encoded in $S_\mathrm{EW}$. Isospin-breaking (IB) enters $R_\mathrm{IB}(s)$, which can be written as
\begin{equation}
R_\mathrm{IB}(s)=\frac{FSR(s)}{G_\mathrm{EM}(s)}\frac{\beta^3_{\pi^+\pi^-}(s)}{\beta^3_{\pi^0\pi^-}(s)}\Bigg|\frac{F_V(s)}{f_+(s)}\Bigg|^2\,,
\end{equation}
where $G_\mathrm{EM}(s)$ is the long-distance electromagnetic radiative corrections factor \cite{Cirigliano:2001er, Cirigliano:2002pv}, that is challenging, as well as the ratio of the neutral-to-charged pion form factors.
 The different results obtained either way can be interpreted as new physics (with Wilson coefficients $\epsilon_i$) affecting tau decays \cite{Cirigliano:2018dyk,Cirigliano:2021yto}
\begin{equation}
\frac{a_\mu^\tau-a_\mu^{e^+e^-}}{2a_\mu^{e^+e^-}}=\epsilon_L^{d\tau}-\epsilon_L^{de}+\epsilon_R^{d\tau}-\epsilon_R^{de}+1.7\epsilon_T^{d\tau}\,,
\end{equation}
whose implications have also been studied in refs.~\cite{Miranda:2018cpf,Gonzalez-Solis:2020jlh}.
The very precise BMW coll. evaluation \cite{Borsanyi:2020mff} also challenged the $e^+e^-$ data-driven $a_\mu^{HVP}$. Their result would imply only a $1.7\sigma$ deviation with respect to the world average.
Furthermore, the recent CMD-3 measurement \cite{CMD-3:2023alj, CMD-3:2023rfe} of the $e^+e^-$ cross-section conflicts severely with previous data, particularly with KLOE's \cite{KLOE-2:2017fda} (not so much with the other very accurate measurement, from BaBar \cite{BaBar:2012bdw}). CMD-3 alone would imply an $a_\mu$ SM prediction in agreement with the measurement within one sigma, as advocated by the recent mixed lattice--data driven evaluation of ref.~\cite{Boccaletti:2024guq}.

\section{Long-distance radiative corrections}
The $G_\mathrm{EM}$ factor was originally studied by Cirigliano \textit{et al.} in refs.~\cite{Cirigliano:2001er, Cirigliano:2002pv}, where it was computed in Resonance Chiral Theory ($R\chi T$)~\cite{Ecker:1988te, Ecker:1989yg}~\footnote{$R\chi T$ has been used to successfully evaluate other contributions to $a_\mu$ \cite{Roig:2014uja, Guevara:2018rhj, Roig:2019reh, Qin:2020udp, Wang:2023njt, Qin:2024ulb,Estrada:2024cfy}.}, including those operators that saturate the chiral low-energy constants up to $\mathcal{O}(p^4)$ \cite{Weinberg:1978kz, Gasser:1983yg, Gasser:1984gg}. A recalculation of this factor was performed by Flores-B\'aez \textit{et al.} \cite{Flores-Baez:2006yiq} using a vector meson dominance (VMD) model. Both results agreed but for the contribution due to the diagrams with a $\rho$-$\omega$-$\pi$ vertex in VMD (which appears at $\mathcal{O}(p^6)$ in $R\chi T$). In ref.~\cite{Miranda:2020wdg} we extended the $R\chi T$ computation including operators contributing up to $\mathcal{O}(p^6)$ in the chiral expansion and confirmed the important r\^ole played by the odd-intrinsic parity sector contributions to the $G_\mathrm{EM}(s)$. We also considered either the short-distance constraints on the $R\chi T$ operators rendering well-behaved two-point correlators \cite{Ecker:1988te, Ecker:1989yg} or extending this to three point-functions \cite{Cirigliano:2006hb, Kampf:2011ty, Roig:2013baa}. Our results agree with other tau-based determinations \cite{Maltman:2005yk,Maltman:2005qq,Davier:2010fmf,Davier:2010nc,Davier:2013sfa,Bruno:2018ono,Narison:2023srj,Esparza-Arellano:2023dps}, with our IB contributions to $a_\mu^\mathrm{HVP,LO}|_{\pi\pi}$ in the range $[-20.52,-6.96]\times10^{-10}$ at $68\%$ confidence level (we are more precise in the recent \cite{Castro:2024prg}).

\section{\texorpdfstring{$\boldsymbol{a_\mu^\mathrm{HVP}}$}{Lg} evaluations}
\begin{figure}[ht]
    \centering
    \includegraphics[width=8cm]{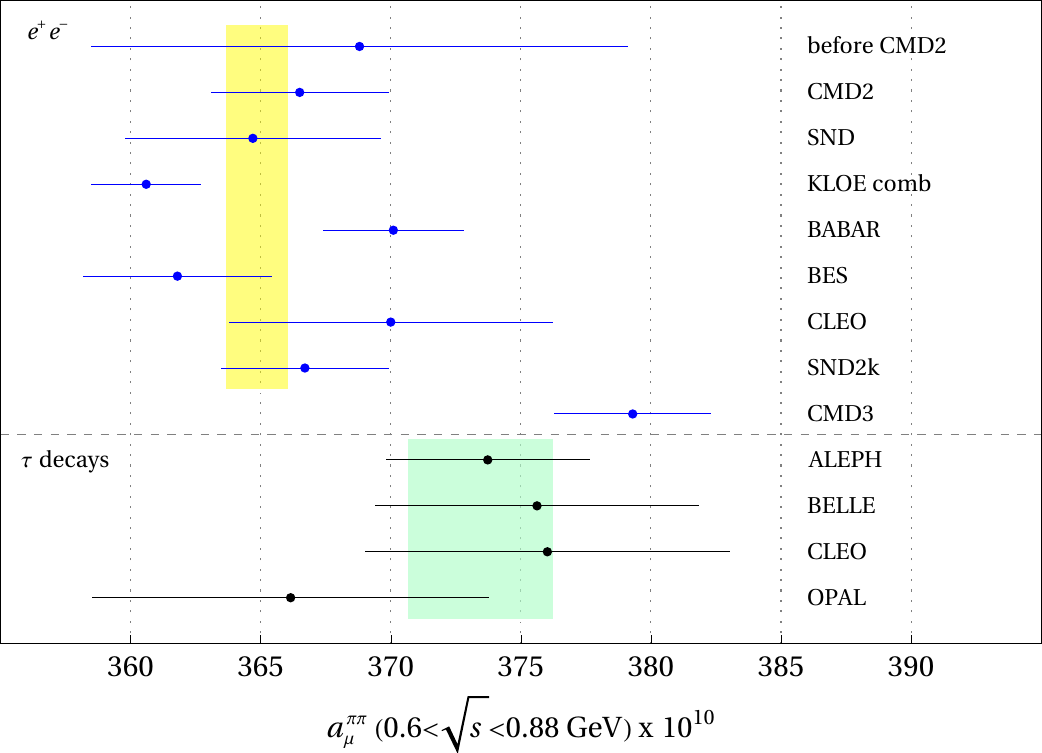}
    \caption{The $\pi\pi(\gamma)$ contribution to $a_\mu^{\text{HVP, LO}}$ around the $\rho$ peak, obtained from the $e^+e^-\to\pi^+\pi^-(\gamma)$ cross section (top) and di-pion $\tau$ decays (bottom).}
    \label{fig:amu_eetau}
\end{figure}

The tension between the different sets of $e^+e^-$ data can be appreciated in 
Fig.~\ref{fig:amu_eetau}, showing the $\pi\pi$ contribution to $a_\mu^\text{HVP, LO}$ around the $\rho$ peak, found using either $\sigma(e^+e^-\to\mathrm{hadrons})$ (top part of the plot, the average in yellow excludes the CMD-3 point, to emphasize its impact) or the $\tau^-\to\pi^-\pi^0\nu_\tau$ spectrum (bottom of the figure, with mean in green which agrees with CMD-3). Clearly, tau data yields a larger value, by $\sim10\times10^{-10}$.

In ref.~\cite{Masjuan:2023qsp} we applied these results to the window quantities introduced in Ref.~\cite{RBC:2018dos}, and used in \cite{Colangelo:2022vok}. We recall that the different contributions of these windows, short-distance ($SD$), intermediate ($int$) and long-distance ($LD$), to $a_\mu^\text{HVP}$ scale as $\sim1:10:25$, respectively, so that the relative accuracy needed varies substantially between them.\\

\begin{figure}[ht!]
    \centering
    \includegraphics[width=8.2cm]{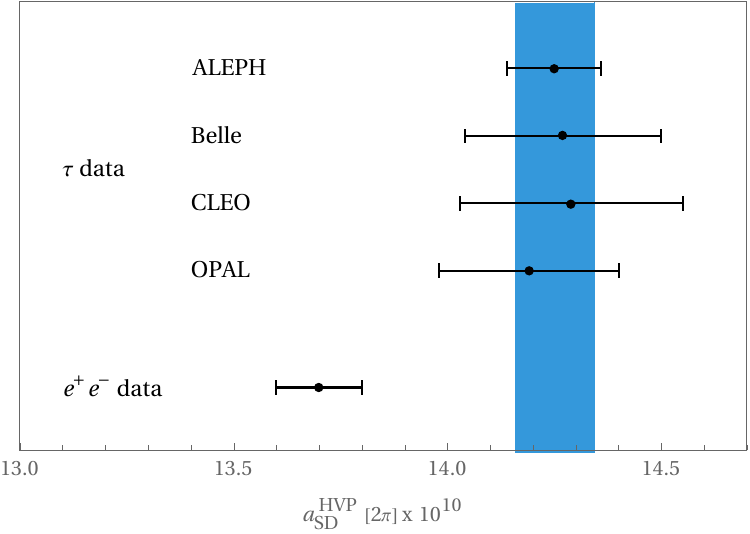}
    \includegraphics[width=8.0cm]{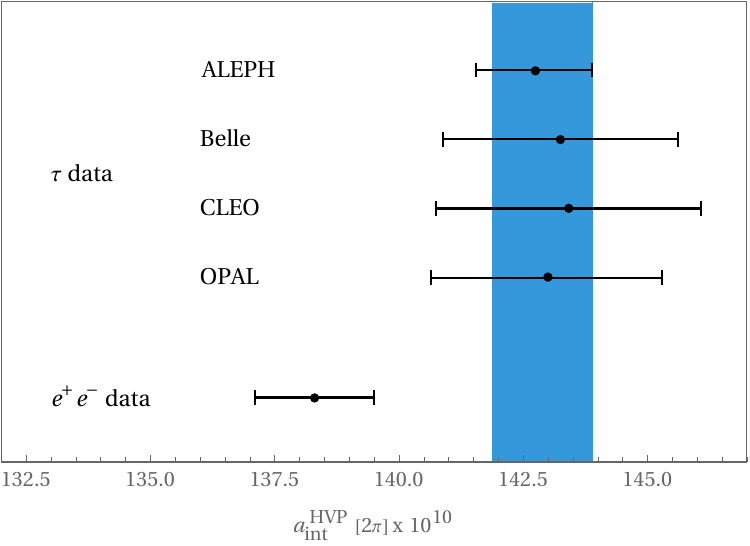}
    \includegraphics[width=8.2cm]{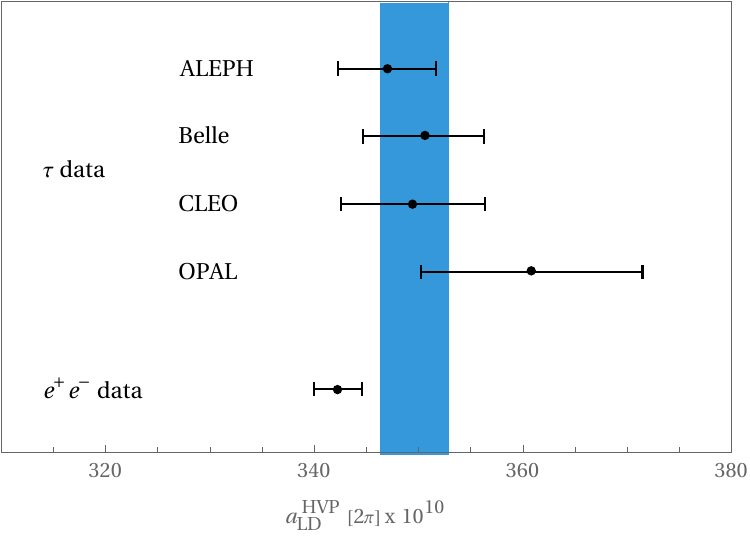}
    \caption{Window quantities ($SD$ top, $int$ medium, and $LD$ bottom) for the $2\pi$ contribution below $1.0\,\text{GeV}$ to $a_\mu^{\mathrm{HVP}}$, corresponding to our reference results. 
 The $\tau$ data mean is shown in blue, with the $e^+e^-$ result from \cite{Colangelo:2022vok}. }
    \label{fig:ChPTOp4}
\end{figure}

 Our most important results for the three different window contributions to $a_\mu^{\mathrm{HVP}}$ are represented in Fig.~\ref{fig:ChPTOp4}, where the different $\tau$ measurements \cite{ALEPH:2005qgp, Belle:2008xpe, CLEO:1999dln, OPAL:1998rrm} agree remarkably. In the $SD$ and $int$ windows, $e^+e^-$ (from Ref.~\cite{Colangelo:2022vok}) and $\tau$ data-based results are at variance.\\

\begin{figure}[ht!]
    \centering
    \includegraphics[width=8.3cm]{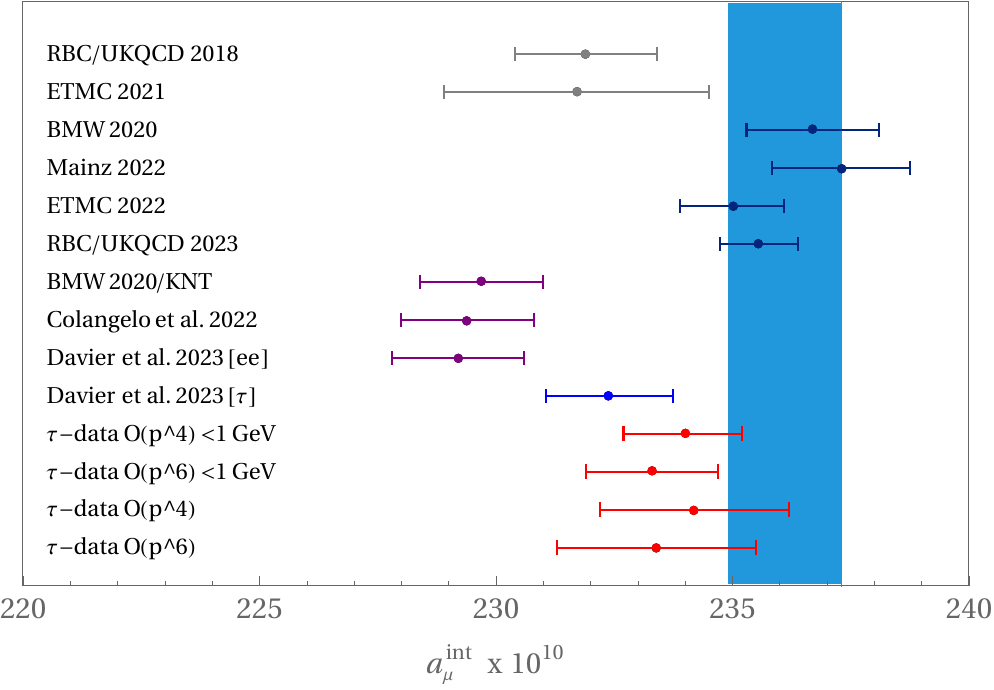}
    \caption{Comparison of the total intermediate window contribution to $a_{\mu}^{\text{HVP, LO}}$ according to lattice QCD, $e^+e^-$ and $\tau$ data-driven evaluations. The blue band is the weighted average of the lattice results excluding those superseded, RBC/UKQCD 2018~\cite{RBC:2018dos} and ETMC 2021~\cite{Giusti:2021dvd} (by refs.~\cite{RBC:2023pvn} and \cite{ExtendedTwistedMass:2022jpw}, respectively).}
    \label{fig:lattice_results}
\end{figure}

Our $\tau$-based $\pi\pi$ contribution to $a_\mu^\text{HVP, LO}$ is complemented with that from the other modes, to confront it directly with the full evaluations. We considered two approaches, as detailed in Ref.~\cite{Masjuan:2023qsp}, and the difference between them gave the associated error to this procedure. Thus, we obtained the results displayed in Fig.~\ref{fig:lattice_results}. A clear tendency of $\tau$-based evaluations agreeing with the lattice outcomes \cite{Giusti:2021dvd,RBC:2023pvn,Ce:2022kxy,ExtendedTwistedMass:2022jpw} is exhibited, whereas the $e^+e^-$ ones differ clearly with both (our results were corroborated by ref.~\cite{Davier:2023fpl}). This discrepancy is almost entirely due to the light-quark connected contribution, dominated by the $\pi\pi$ channel (it is $\sim 81\%$) \cite{Benton:2023dci,Benton:2024kwp}.

\section{Conclusions}

There is a global effort in improving the evaluation of the hadronic contributions to $a_\mu$. Specifically, dedicated studies to improve the HVP part --which dominates the SM uncertainty-- from lattice, dispersion relations, data-driven methods, improved $e^+e^-$ data and Monte Carlo generators for the low-energy hadron cross-section \cite{WorkingGrouponRadiativeCorrections:2010bjp} are being undertaken.

Through the years, the tau-data driven computation has always been approximately $[2,2.5]\sigma$ away from the experimental average, while the tension with $e^+e^-$ data was systematically larger than three sigmas. The most recent lattice QCD results by the Mainz/CLS, ETMC, RBC/UKQCD Colls. agree remarkably with BMW in the intermediate window. It is then of utmost importance than another lattice computation reaches a comparable accuracy to BMW in the long-distance window.

We showed that tau based results are compatible with the lattice evaluations in the intermediate window, while the $e^+e^-$ data are in tension with both. This puzzle deserves further scrutiny.

\acknowledgments
We gladly acknowledge the QNP24 organizing committees. We thank financial support through the Ministerio de Ciencia e Innovación under Grant No. PID2020–112965 GB-I00, by the Departament de Recerca i Universitats from Generalitat de Catalunya to the Grup de Recerca ``Grup de Física Teòrica UAB/IFAE'' (Codi: 2021 SGR 00649) and partial Conahcyt support, from project CB2023-2024-3226.

\end{document}